\documentclass[aps,twocolumn,showpacs]{revtex4}
\usepackage{graphicx}
\def\bsigma{\mbox{\boldmath $\sigma$}}
\def\bSigma{\mbox{\boldmath $\Sigma$}}

\def\bOmega{\mbox{\boldmath $\Omega$}}

\begin{document}
\title{Intrinsic Spin Hall Effect: Topological Transitions in
Two-Dimensional Systems}
\author{O. E. Raichev}
\email{raichev@isp.kiev.ua}
\affiliation{Institute of Semiconductor Physics,
National Academy of Sciences of Ukraine,
Prospekt Nauki 45, 03028, Kiev, Ukraine}

\begin{abstract}
The spin-Hall conductivity in spatially-homogeneous two-dimensional electron systems
described by the spin-orbit Hamiltonian $\hbar \bOmega_{\bf p}\cdot
\hat{\bsigma}$ is presented as a sum of the universal part $Me/8 \pi \hbar$
determined by the Berry phase $\Phi=M \pi$ ($M$ is an odd integer, the
winding number of the vector $\bOmega_{\bf p}$) and a non-universal part which
vanishes under certain conditions determined by the analytical properties
of $\bOmega_{\bf p}$. The analysis reveals a rich and complicated behavior of the
spin-Hall conductivity which is relevant to both electron and hole states in quantum
wells and can be detected in experiments.
\end{abstract}


\maketitle

Owing to the spin-orbit interaction (SOI), an electric field applied
along two-dimensional (2D) electron layers can generate transverse spin
currents in the absence of external magnetic fields. This phenomenon,
known as the spin Hall effect [1], is at the focus of attention in
modern physics. The presence of SOI terms in the Hamiltonian of free
electrons leads to the intrinsic spin-Hall conductivity expressed in the
universal units $e/\hbar$ in the case of weak disorder. The original
theoretical proposal [2] of the universal intrinsic spin Hall effect has
been based on the Rashba Hamiltonian [3] describing the linear in 2D
momentum ${\bf p}=(p_x,p_y)$ spin-orbit coupling due to structural
inversion asymmetry. However, numerous theoretical calculations [4-8]
have proved the absence of static intrinsic spin currents for
this case. As follows from the equation of motion for the spin
density operator [9], this statement is applicable to any electron
system described by a ${\bf p}$-linear SOI Hamiltonian. The situation
is quite different in the case of 2D hole systems described by the
effective ${\bf p}$-cubic SOI Hamiltonian [10] $h^{ (3)}_{\bf p}
= \hbar \kappa (\hat{\sigma}_{+} p_-^3 + \hat{\sigma}_{-} p_+^3)$,
where $\hat{\sigma}_{\pm}=(\hat{\sigma}_x \pm i \hat{\sigma}_y)/2$,
$\hat{\sigma}_{\alpha}$ are the Pauli matrices, and $p_{\pm}=p_x \pm i p_y$.
Theoretical studies [11-15] based upon this Hamiltonian have confirmed the
existence of the intrinsic spin Hall effect. The experimentally observed
spin Hall effect in 2D hole systems [16,17] is likely of the intrinsic origin.

What makes the systems described by the Hamiltonian $h^{ (3)}_{\bf p}$ so
different from the systems described by {\bf p}-linear SOI Hamiltonians?
It is the dependence of SOI on the angle $\varphi$ of the 2D momentum. This
dependence is characterized by the odd integers known as the winding numbers
(WN), which are equal to $\pm 3$ and $\pm 1$ for the ${\bf p}$-cubic and
${\bf p}$-linear Hamiltonians considered above. In 2D hole systems, owing
to the increased WN, the conservation of the spin density is no longer reduced
to the requirement of zero spin currents, so the intrinsic spin Hall effect
exists. The role of WN in spin response can be also emphasized by considering
their influence on the collision-mediated spin-charge coupling term known from
the Kubo formalism as the vertex correction [4]. If the scattering is symmetric
(caused by the short-range potential), the vertex correction for the WN $\pm 3$
is zero, since it is given by the angular average of the product of the
charge current operator by the SOI Hamiltonian. In contrast, for ${\bf p}$-linear
SOI the vertex correction is always nonzero and leads to nonexistence of spin
currents.

It is important to realize that the consideration of SOI Hamiltonians containing
the terms with WN either $\pm 1$ or $\pm 3$ is not sufficient for description of
the spin response in 2D systems. The coexistence of SOI terms with WN $\pm 1$ and
$\pm 3$ in semiconductor quantum wells is rather a rule than an exception. For
example, this is the case of conduction-band electrons in the quantum wells made
of noncentrosymmetric semiconductors [18] at high electron densities, when both
${\bf p}$-linear and ${\bf p}$-cubic Dresselhaus terms are important. The aim of
this Letter is to find out the general properties of the intrinsic spin currents
for the systems described by the SOI Hamiltonians containing an arbitrary mixture
of terms with different WN and to establish relevance of such a consideration to
both electron and hole states in quantum wells.

The starting point is the free-electron Hamiltonian in the momentum representation:
\begin{equation}
\hat{H}_{\bf p} = \varepsilon_p + \hat{h}_{\bf p},~~~
\hat{h}_{\bf p}=\hbar \bOmega_{\bf p} \cdot \hat{\bsigma},
\end{equation}
where $\varepsilon_p$ is the kinetic energy (isotropic but not
necessarily parabolic) and $\bOmega_{\bf p}$ is an {\em arbitrary}
vector antisymmetric in momentum. The $2 \times 2$ matrix SOI term
$\hat{h}_{\bf p}$ describes both 2D electrons and 2D holes (since the 4-fold
degeneracy of the $\Gamma_8$ valence band is lifted in quantum wells, the
2D holes are quasiparticles with two spin states). The calculations are
based on the quantum kinetic equation for the Wigner distribution function
[19] which is a $2 \times 2$ matrix over the spin indices.
Searching for the linear response to the applied electric field ${\bf E}$
in the stationary and spatially homogeneous case, one can write the
distribution function in the form $\hat{f}^{(eq)
}_{\mathbf{p}}+\hat{f}_{\mathbf{p}}$, where $\hat{f}_{\mathbf{p}}$ is the
non-equilibrium part satisfying the linearized kinetic equation
\begin{equation}
\frac{i}{\hbar} \left[ \hat{h}_{\mathbf{p}},\hat{f}_{\mathbf{p}}\right]
+ e {\bf E} \cdot \frac{\partial \hat{f}^{(eq)}_{\mathbf{p}}}{\partial {\bf p}} =
\widehat{J} ( \hat{f}| \mathbf{p} ).
\end{equation}
The collision integral $\widehat{J}$ describes the elastic scattering,
and the spin-orbit corrections [1] to the scattering potential are neglected.
Considering this integral in the Markovian approximation and assuming
that $\hbar \Omega_{\bf p} \equiv \hbar |{\bOmega_{\bf p}}|$ is small in
comparison to the mean kinetic energy, one can expand $\widehat{J}$ in series
of $\Omega_{\bf p}$ [14,19]. Using the spin-vector representation $\hat{f}_{\bf p}=
{\rm f}^0_{\bf p}+ \hat{\bsigma} \cdot {\rm {\bf f}}_{\bf p}$, one gets
\begin{equation}
-2 [ \bOmega_{\bf p} \times {\rm {\bf f}}_{\bf p}] +
{\bf A}_{\bf p} = \frac{m_p}{\hbar^3} \int_0^{2 \pi} \! \!
\frac{d \varphi'}{2 \pi} w_{|{\bf p}-{\bf p}'|} ({\rm {\bf f}}_{\bf p'}-
{\rm {\bf f}}_{\bf p}),
\end{equation}
where $w_q$ is the Fourier transform of the correlator
of the scattering potential, $|{\bf p}'|=|{\bf p}|$ is assumed, and
$\varphi'$ is the angle of the vector ${\bf p}'$. Next, ${\bf A}_{\bf p}$
is a vector proportional to ${\bf E}$, and $m_p \equiv
\frac{1}{2} (\partial p^2/\partial\varepsilon_p)$ is the
$p$-dependent effective mass as it enters the expression for the group velocity,
${\bf v}_{\bf p}= \partial \varepsilon_p/\partial {\bf p}={\bf p}/m_p$.
Analytical solution of Eq. (3) is possible for short-range scattering potential,
when $w_{|{\bf p}-{\bf p}'|} \simeq w$. Then the right-hand side of Eq. (3) is
written as $\nu_p (\overline{ {\rm {\bf f}}}_{\bf p} - {\rm {\bf f}}_{
\bf p})$, where $\nu_p=m_pw/\hbar^3$ is the scattering rate and
the line over a function denotes the angular averaging. Also,
\begin{equation}
\frac{A^{\alpha}_{\bf p}}{\hbar e} = {\bf E} \cdot \!
\frac{\partial ({\bf v}_{\bf p} \Omega^{\alpha}_{\bf p} -
\overline{{\bf v}_{\bf p} \Omega^{\alpha}_{\bf p}})
f'_{\varepsilon_p}}{\partial \varepsilon_p} + [{\bf E} \times {\bf n}]
\cdot \frac{f'_{\varepsilon_p}}{p^2} \frac{\partial {\bf p} \Omega^{\alpha}_{\bf p}
}{\partial \varphi},
\end{equation}
where $f'_{\varepsilon_p} \equiv \partial f_{\varepsilon_p}/ \partial
\varepsilon_p$ is the derivative of the Fermi distribution function
$f_{\varepsilon}$, and ${\bf n}$ is the unit vector normal to the
quantum well plane. Notice the property $\overline{A^{\alpha}_{\bf p}}=0$.

Solution of Eq. (3) determines the non-equilibrium spin current density
${\bf q}_{\gamma}= \frac{1}{2} \int \frac{d {\bf p}}{(2 \pi \hbar)^2}
{\rm Tr}( \{ \hat{\bsigma} , \hat{u}_{\gamma}({\bf p}) \}
\hat{f}_{\bf p} )$, where $\hat{u}_{\gamma}({\bf p})=\partial
(\varepsilon_p+\hat{h}_{\bf p})/\partial p_{\gamma}$ is the group velocity
in the presence of spin-orbit interaction, $\{~,~\}$ denotes the
symmetrized matrix product, and ${\rm Tr}$ is the matrix trace. The spin
conductivity is introduced according to ${\bf q}_{\gamma}=
\bSigma_{\gamma \beta} E_{\beta}$. Based on Eqs. (3) and (4),
\begin{eqnarray}
\bSigma_{\gamma \beta}= -\frac{e}{8 \pi \hbar} \int d \varepsilon_p ~ f'_{\varepsilon_p}
\left( {\bf T}^{\gamma \beta}_p - [\overline{ {\bf P}^{\gamma}_{\bf p}}  \times
{\bf Q}^{\beta}_p] \right).
\end{eqnarray}
The vector-functions standing here are defined as angular averages:
${\bf T}^{\gamma \beta}_p=2 \overline{ \left[ {\bf P}^{\gamma}_p \times
(\partial \bOmega_{\bf p}/\partial p_{\beta}) \right] }$,
${\bf P}^{\gamma}_{\bf p}=p_{\gamma} \bOmega_{\bf p}/\Delta^2_{\bf p}$,
$\Delta^2_{\bf p}=\Omega^2_{\bf p} + \nu^2_p/4$, and ${\bf Q}^{\beta}_p=
2 [\overline{\widehat{R}}_p]^{-1} \overline{ \widehat{R}_p (\partial
\bOmega_{\bf p}/\partial p_{\beta}) }$, where $\widehat{R}_p$ is a symmetric
matrix with elements $R^{\alpha \beta}_p= (\Omega^2_{\bf p} \delta_{\alpha
\beta}-\Omega^{\alpha}_{\bf p} \Omega^{\beta}_{\bf p})/\Delta^2_{\bf p}$.
One can find also the induced spin density: ${\bf s}= \frac{1}{2} \int
\frac{d {\bf p}}{(2 \pi \hbar)^2} {\rm Tr}( \hat{\bsigma} \hat{f}_{\bf p})=
(e \hbar^2/4 \pi w) \int d \varepsilon_p f'_{\varepsilon_p} {\bf Q}^{\beta}_p
E_{\beta}$. The limit of low temperature [20] is described
by the substitution $f'_{\varepsilon_p}=-\delta(\varepsilon_p-
\varepsilon_{ F})$, so the spin conductivity tensor is expressed
directly through the vector-functions taken at the Fermi surface
$\varepsilon_p=\varepsilon_{p_F}=\varepsilon_{ F}$.

Equation (5) is valid for arbitrary $\bOmega_{\bf p}$. In the
quantum wells grown along [001] direction in
cubic crystals of zinc-blende type, the $C_{2v}$ point group
symmetry implies
\begin{equation}
\bOmega_{\bf p}=(\Omega^{x}_{\bf p}, \Omega^{y}_{\bf p},0),~~
\Omega^{x}_{p, -\pi/4 + \varphi} = \Omega^{y}_{p, -\pi/4 - \varphi},
\end{equation}
where the polar coordinate representation ${\bf p} \equiv (p, \varphi)$
is used. Then ${\bf T}^{\gamma \beta}_p=(0,0,T^{\gamma \beta}_p)$,
$\overline{ {\bf P}^{\gamma}_{\bf p}}=(P_p^{x \gamma},P_p^{y \gamma},0)$,
and ${\bf Q}^{\beta}_p=(Q_p^{x \beta},Q_p^{y \beta},0)$, where
$T^{xx}_p=-T^{yy}_p$, $T^{xy}_p=-T^{yx}_p$, $P^{xx}_p=-P^{yy}_p$,
$P^{xy}_p=-P^{yx}_p$, $Q^{xx}_p=-Q^{yy}_p$, and $Q^{xy}_p=-Q^{yx}_p$.
The spin currents exist only for $z$-spins, $\bSigma_{\gamma \beta}=
(0,0,\Sigma_{\gamma \beta})$, and there are two independent components
$\Sigma_{xy}=-\Sigma_{yx} \equiv \Sigma_{H}$ and $\Sigma_{xx}=-
\Sigma_{yy}$ describing spin-Hall and spin-diagonal currents,
respectively. The function $T^{xy}_p$ entering $\Sigma_{H}$
can be written as
\begin{equation}
T^{xy}_p= \int_0^{2 \pi} \! \! \frac{d \varphi}{2 \pi \Delta^2_{p,\varphi}}
\left( \Omega^{x}_{p,\varphi} \frac{\partial \Omega^y_{p,\varphi}}{
\partial \varphi} - \Omega^{y}_{p,\varphi}
\frac{\partial \Omega^x_{p,\varphi}}{\partial \varphi} \right).
\end{equation}

In the case of zero temperature, using the notations
$T^{xy} \equiv T^{xy}_{p_F}$ and $\bOmega_{\varphi} \equiv
\bOmega_{p_F,\varphi}$, it is convenient to write
\begin{equation}
\Sigma_{H}= \frac{e}{8 \pi \hbar} T^{xy} + \delta \Sigma_{H},
\end{equation}
where $\delta \Sigma_{H}$ expresses the contribution of the second
term in Eq. (5). In the collisionless limit, the formal integration
in Eq. (7) leads to
\begin{equation}
T^{xy}=\frac{\Phi}{\pi},~~\Phi= \frac{1}{2}
\oint d \arg[\Omega^{+}(z)]=\pi (N_0-N_{\infty}),
\end{equation}
where $\Omega^{+}(z)=\Omega^x_{\varphi} + i \Omega^y_{\varphi}$ is a
function of the complex variable $z=e^{i \varphi}$, and the contour
of integration in the complex plane is the circle of unit radius,
$|z|=1$. Next, $N_0$ and $N_{\infty}$ are the numbers of zeros and
poles of $\Omega^{+}(z)$ inside this circle (it is assumed that
$\Omega^{+}(z)$ does not have branch points). Using the conventional
definitions (see [21] and references therein) it is easy to identify $\Phi$
with the Berry phase in the momentum space. In the WN representation,
the function $\Omega^{+}(z)$ is a polynomial containing {\em odd}
powers of $z$, in the general case, from $z^{-N}$ to $z^N$, assuming
that the highest WN involved in $\bOmega_{\bf p}$ is $N$. Then $N_{\infty}=L$,
where $L \leq N$ is an odd integer (the order of the multiple pole
at $z=0$), while $N_0$ takes {\em even} values from $0$ to $L+M$, where
$-L \leq M \leq N$ depending on the SOI parameters. Therefore, if
$\Omega_{\bf p}$ contains an arbitrary mixture of terms with different
WN up to $N$, the spin-Hall conductivity is
\begin{equation}
\Sigma_{H}= \frac{M e}{8 \pi \hbar} + \delta \Sigma_{H},~~~|M|=1,3, \ldots, N,
\end{equation}
where $M=N_0-N_{\infty}$ is the {\em acting} WN, which describes the
actual winding of the vector $\bOmega_{\bf p}$ as ${\bf p}$ goes around the
Fermi surface, and can be found, in each concrete case, from the simple analysis
explained above. The corresponding Berry phase is $\pi M$. The spin-Hall conductivity
changes abruptly when the functions $\Omega^x_{\varphi}$ and $\Omega^y_{
\varphi}$ go through zero simultaneously at certain angles $\varphi$. In
other words, each time when the SOI parameters are adjusted in such a
way that the spin splitting $2 \hbar \Omega_{\bf p}$ at the Fermi
surface becomes zero at certain ${\bf p}$, a topological transition occurs: the Berry
phase changes by $\pm 2 \pi$. For the Hamiltonians with $N=1$ including both Rashba and
Dresselhaus (linear) terms, this effect has been studied in the Berry phase approach
in Refs. 21-23. In this particular case, however, the first term in Eq. (10)
is exactly compensated by the second term, and $\Sigma_{H}=0$. Therefore, the
topological transitions essentially require the SOI with WN greater than unity.

The result (10) is exact in the collisionless limit and can be viewed as a
quantization of the spin-Hall conductivity in terms of the WN. In general, this
quantization does not occur in integer numbers of $e/4 \pi \hbar$, because
$\delta \Sigma_{H}$ is also a discontinuous function of SOI parameters and
undergoes abrupt changes together with the first term in Eq. (10). To show this, it
is sufficient to represent $P^{\alpha \gamma}$ as combinations of the integrals
$\oint d z [\Omega^{+}(z)]^{-1}$, $\oint d z [z^2 \Omega^{+}(z)]^{-1}$, and
complex conjugate terms. It is important that such a representation allows one to find
the general conditions for vanishing $\delta \Sigma_{H}$: this takes place when either
a) all zeros of $\Omega^{+}(z)$ are inside the circle $|z|=1$ and $N_0-N_{\infty}
\geq 3$ or b) the order of the multiple pole at $z=0$ is $L \geq 3$ and all
zeros of $\Omega^{+}(z)$ (if present) are outside the circle $|z|=1$.
In particular, this means that if the highest WN involved in $\bOmega_{\bf p}$
is, in the same time, the acting WN ($M=N$ or $M=-N$ at $N \neq 1$), the spin-Hall
conductivity stays at the universal value $M e/8 \pi \hbar$ without regard to
the SOI parameters. If $N \geq 5$, $\Sigma_{H}$ can take universal values
from $\pm 3 e/8 \pi \hbar$ to $\pm N e/8 \pi \hbar$.

\begin{figure}[ht]
\begin{center}
\includegraphics[scale=0.4]{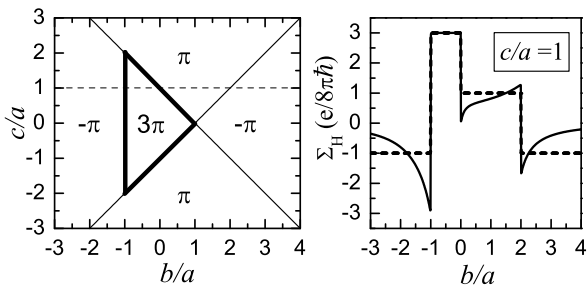}
\end{center}
\addvspace{-0.7 cm}\caption{Left: Phase diagram
for the SOI of Eq. (11) at $u_p=0$. The regions of fixed
Berry phase $\Phi$ (indicated) are separated by the lines
of topological transitions (solid). Right: Spin-Hall
conductivity $\Sigma_{H}$ (solid) and its universal part
(dash) as functions of $b/a$ at $c/a=1$. It is assumed
that $\partial \ln |b_p|/\partial \ln p=\partial \ln |c_p|/\partial
\ln p=\!1$ and $\partial \ln |a_p|/\partial \ln p=\!3$.}
\end{figure}

The most general form of $\bOmega_{\bf p}$ including WN $\pm 1$
and $\pm 3$ for [001]-grown quantum wells is
\begin{eqnarray}
\Omega^x_{\bf p}=  c_p \sin\varphi- b_p \cos\varphi -u_p \sin 3 \varphi
- a_p \cos 3\varphi, \nonumber \\
\Omega^y_{\bf p}= -c_p \cos\varphi+ b_p \sin\varphi -u_p \cos 3 \varphi
- a_p \sin 3\varphi.
\end{eqnarray}
This form describes both electron and hole states. For conduction-band
electrons, there are the Rashba ($c_p$) and the Dresselhaus ($b_p$) terms, while
the $a_p$-term exists because of the ${\bf p}$-cubic Dresselhaus contribution.
The $u_p$-term can be attributed to higher-order invariants allowed by
symmetry. For holes in the ground-state subband, the $a_p$- and $b_p$-terms
exist due to the structural inversion asymmetry. The term containing $a_p
\propto p^3$ is the one considered in the theory of the spin Hall effect for
holes, this term is derived [10] from the isotropic Luttinger Hamiltonian.
The anisotropy of the Luttinger Hamiltonian, described by the parameter
$\mu=(\gamma_2-\gamma_3)/(\gamma_2+\gamma_3)$, where $\gamma_i$ are the Luttinger
parameters in their usual notations, leads to the $b_p$-term with $b_p=\mu a_p$.
Next, the $c_p$- and $u_p$-terms for holes are caused by the bulk inversion
asymmetry [24]. The $c_p$-term includes the contribution $\alpha_h p$
proportional to $p$ [24,25], which should dominate at low hole densities.
In the general case, especially when the structural asymmetry is weak, an
adequate description of hole states should include all terms in Eq. (11).

\begin{figure}[ht]
\begin{center}
\includegraphics[scale=0.4]{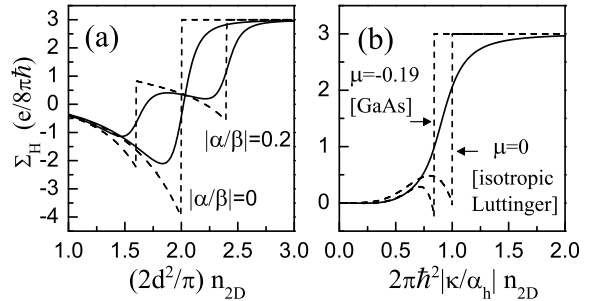}
\end{center}
\addvspace{-0.7 cm}\caption{Spin-Hall conductivity as a function of
density in electron (a) and hole (b) systems. The dashed lines correspond
to the collisionless approximation, $\nu=0$. The solid lines are plotted for
$\nu=0.2 \beta (\pi \hbar/d)$ (a) and for $\nu=0.5 |\alpha_h^3/\kappa|^{1/2}$ (b).}
\end{figure}

The simplest case of the SOI with combined WN described by Eq. (11)
is realized when $c_p=u_p=0$. One finds the analytical expression
\begin{equation}
\Sigma_{H}=\frac{e}{8 \pi \hbar} \frac{a^2+b^2-r^2}{2 b^2 r^2}
\left[(3 -\eta)(a^2-b^2) + \eta r^2 \right],
\end{equation}
where $r_p^2= \sqrt{[(a_p+b_p)^2 \! + \! \nu_p^2/4][(a_p-b_p)^2 \!
+ \! \nu_p^2/4]} - \! \nu_p^2/4$, $\eta_p = 1/2 + (1/4)
\partial \ln|b_p/a_p|/\partial \ln p$, and all coefficients are taken
at $p=p_{ F}$. According to the Berry phase analysis, $\Sigma_{H}=3e/8 \pi
\hbar$ at $|a|>|b|$ in the collisionless limit, while Eq. (12) gives
\begin{equation}
\Sigma_{H}=\frac{3 e}{8 \pi \hbar} \times \left\{ \begin{array}{l} 1, \\
-(1-2 \eta/3 ) (a/b)^2, \end{array} \begin{array}{c} a^2 > b^2 \\
a^2 < b^2 \end{array} \right. .
\end{equation}
In application to conduction-band electrons, when the Dresselhaus
model implies $a_p=\lambda p^3$, $b_p=\beta p-\lambda p^3$, and $\lambda \simeq
\beta (d/2 \pi \hbar)^2$ (for a deep square well
of width $d$), this means that $\Sigma_{H}$ abruptly jumps to the
universal value $3e/8 \pi \hbar$ if the electron density
$n_{ 2D} = p_{ F}^2/2 \pi \hbar^2$ increases and exceeds $\pi/d^2$.
A similar behavior, though without a qualitative explanation, has been
found in Ref. 26. For holes, $a_p=-\kappa p^3$, $b_p=-\mu
\kappa p^3$, and $|a|>|b|$ since $|\mu|<1$. This means that
$\Sigma_{H}$ of 2D holes in [001]-grown wells is
insensitive to the anisotropy of the Luttinger Hamiltonian
and stays at the universal value for the case of clean hole systems.

\begin{figure}[ht]
\begin{center}
\includegraphics[scale=0.4]{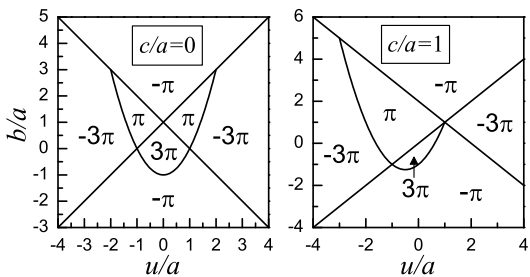}
\end{center}
\addvspace{-0.7 cm}\caption{Phase diagrams for the SOI of Eq. (11).
The Berry phases $\Phi$ for each region are indicated.
The spin-Hall conductivity is $\pm 3e/8 \pi \hbar$ in the
regions with $\Phi=\pm 3 \pi$.}
\end{figure}

If the $c_p$-term is added into consideration, the analysis leads
to the phase diagram shown in Fig. 1. The spin-Hall conductivity
is equal to $3e/8 \pi \hbar$ at $-1 < b/a < 1-|c/a|$, in the region inside
the bold triangle in Fig. 1. There are 5 regions, and
several topological transitions can take place as the parameters
are varied. To demonstrate a possibility of their experimental
observation, one should put $c_p=\alpha p$ for electrons and $c_p=\alpha_h p$
for holes. The Rashba coefficient $\alpha$ is determined by structural
asymmetry, while $\alpha_h \simeq 2 \delta \alpha_v/(\hbar d)^2$ [24], where
$\delta \alpha_v=-0.035$ eV nm$^3$ for GaAs. The results of calculations
are shown in Fig. 2. For electrons, $\Sigma_H$ is plotted as a function of
the dimensionless parameter $2 d^2 n_{ 2D}/\pi= (p_{ F} d/\pi \hbar)^2$
in the range $p_{ F} < \sqrt{3} \pi \hbar/d$, when only the lowest electron
subband in the deep square well is populated. If Rashba coupling is nonzero,
this dependence has two jumps and the region of universal behavior is shifted
towards higher densities. If $|\alpha/\beta|$ exceeds 1, $\Sigma_H$
becomes considerably suppressed in the chosen density range.
For holes, it is convenient to use the dimensionless
units $2 \pi \hbar^2 |\kappa/\alpha_h| n_{ 2D}=|a/c|$. The transition takes place at
$|a/c|=1/(1-\mu)$. Estimating $\hbar^4 \kappa \sim 0.1$ eV nm$^3$ from the data of Ref. 17
and assuming $d \simeq 5$ nm, one finds that this condition corresponds to $n_{ 2D}
\sim 5 \times 10^{11}$ cm$^{-2}$, so the transition occurs at a reasonable density
and can be observed experimentally. Instead of varying $n_{ 2D}$, it is possible
to change $\alpha$ for electrons and $\kappa$ for holes by biasing the structure.

Finally, after adding the $u_p$-term the phase diagram becomes
more complicated, it is described in terms of three variables,
$c/a$, $b/a$, and $u/a$. Figure 3 shows two sections of
this three-dimensional phase diagram, which demonstrate coexistence
of the regions with $\Phi=3 \pi$ and $\Phi=-3 \pi$, and a possibility
of transitions between them, when $\Sigma_{H}$ changes by
$3 e/4 \pi \hbar$. The regions of $\Phi=-3 \pi$ exist when $|u/a|>1$.
If $|c/a|>3$, the region of $\Phi=3 \pi$ disappears.

In conclusion, the presence of SOI terms with different angular dependences
and interference of these terms in the spin response makes the physics
of the spin Hall effect more rich than it is usually assumed.
The consideration given above is an attempt to plot a map to this new world,
only part of which has been investigated so far.

\end{document}